\documentclass[10pt,conference]{IEEEtran}
\IEEEoverridecommandlockouts

\usepackage{cite}
\usepackage{amsmath,amssymb,amsfonts}
\usepackage{algorithmic}
\usepackage{graphicx}
\usepackage{textcomp}
\usepackage{xcolor}
\usepackage{multirow}

\def\BibTeX{{\rm B\kern-.05em{\sc i\kern-.025em b}\kern-.08em
    T\kern-.1667em\lower.7ex\hbox{E}\kern-.125emX}}

\usepackage{subcaption}
\usepackage{pbalance}

\makeatletter
\newcommand{\linebreakand}{%
  \end{@IEEEauthorhalign}
  \hfill\mbox{}\par
  \mbox{}\hfill\begin{@IEEEauthorhalign}
}
\makeatother
\newcommand{\bm}[1]{{\mbox{\boldmath $#1$}}}

\begin{document}
 
\title{Hypervisor-based Double Extortion Ransomware Detection Method Using Kitsune Network Features}

\author{%
\IEEEauthorblockN{Manabu Hirano}
\IEEEauthorblockA{\textit{Department of Information and Computer Engineering} \\
\textit{National Institute of Technology, Toyota College}\\
Toyota, Japan \\
hirano@toyota-ct.ac.jp}
\and
\IEEEauthorblockN{Ryotaro Kobayashi}
\IEEEauthorblockA{\textit{Faculty of Informatics} \\
\textit{Kogakuin University}\\
Tokyo, Japan \\
ryo.kobayashi@cc.kogakuin.ac.jp}
}

\maketitle

\begin{abstract}
Double extortion ransomware attacks have become mainstream since many organizations adopt more robust and resilient data backup strategies against conventional crypto-ransomware. This paper presents detailed attack stages, tactics, procedures, and tools used in the double extortion ransomware attacks. We then present a novel detection method using low-level storage and memory behavioral features and network traffic features obtained from a thin hypervisor to establish a defense-in-depth strategy for when attackers compromise OS-level protection. We employed the lightweight \emph{Kitsune} Network Intrusion Detection System (NIDS)'s network feature to detect the data exfiltration phase in double extortion ransomware attacks. Our experimental results showed that the presented method improved by 0.166 in the macro F score of the data exfiltration phase detection rate. Lastly, we discuss the limitations of the presented method and future work.
\end{abstract}

\begin{IEEEkeywords}
ransomware, exfiltration, defense-in-depth, deep learning, behavioral features.
\end{IEEEkeywords}
  
\section{Introduction}\label{sec:intro}

Double extortion ransomware attacks encrypt victim files and exfiltrate them for extortion; the attacker demands a double ransom to restore victim files and stop the victim's sensitive data leakage. As organizations adopt more robust backup strategies for conventional crypto-ransomware attacks, more ransomware attacks focus on data breach attacks. The \emph{Maze} ransomware (2020) was the first known ransomware to exfiltrate sensitive data to blackmail victims into paying ransom \cite{mcintosh2021ransomware}. Since many attackers exfiltrate sensitive data before file encryption \cite{conti-playbook}, detecting the early stage of the double extortion ransomware attacks is crucial for an organization's risk mitigation.

Many ransomware attack detection and prevention methods focus on indicators obtained at the post-infection or destruction phase; for example, crypto-ransomware attacks that perform file enumeration and encryption operations can be detected using specific patterns of Application Programming Interface (API) calls, system calls, Input and Output (I/O), file system operations\cite{begovic2023cryptographic}.
Cen et al., on the other hand, presented a state-of-the-art survey of research on ransomware early detection\cite{cen2024ransomware}. They defined early ransomware detection as a continuous monitoring and identification mechanism that detects ransomware attacks before the \emph{destruction} phase using behavioral and static analysis, anomaly detection, and machine learning. The behavioral and static features of ransomware detection include API call sequences, I/O Request Packets (IRP), network activity, Windows event logs, file entropy, Windows Portable Executable (PE) header, Opcode sequence, and file system activity. Singh and Tripathy presented an Android ransomware detection method at the early stage using MITRE ATT\&CK knowledge base \cite{singh2024s}. They used permissions and APIs invoked by Android ransomware during its execution; they used Android ransomware's tactics specified in MITRE ATT\&CK (e.g., reconnaissance, initial access, privilege escalation, command and control, discovery, lateral movement, exfiltration, and impact) in training deep learning models. Since ransomware attacks consist of many phases, from initial accesses to final impact, detecting the early phase of ransomware attacks as much as possible is crucial.

\subsection{Limitation of prior work and new challenge}

Many state-of-the-art ransomware detection methods employ deep learning models using behavioral indicators obtained from API and system call monitoring, Input and Output Request Packets (IRP) monitoring, file system operation monitoring, network traffic monitoring, etc\cite{begovic2023cryptographic,oz2022survey}. Most of them work on the operating system (OS) layer. However, once attackers gain administrative privileges in the operating system, they can hide their existence using evasion techniques such as process injection and direct kernel object manipulation (DKOM). To mitigate the risk of such OS-level attacks, we presented a hypervisor-based ransomware detection method using low-level storage and memory access patterns \cite{HIRANO-CSR2022, HIRANO2022301314, HIRANO2025104202}; it provides an additional protection layer to the conventional Anti-Virus software and Endpoint Detection and Response (EDR) software executed on the operating system since a hypervisor operates on the higher privilege than operating systems.

Although the previous work \cite{HIRANO-CSR2022, HIRANO2022301314, HIRANO2025104202} can detect ransomware attacks in the final \emph{destruction} phase (i.e., file enumeration and encryption phase) using low-level storage and memory access patterns, it cannot detect the early phase of ransomware attacks, including sensitive data exfiltration, because they did not use network features. This paper presents a novel network monitoring function in the hypervisor layer and the use of the network features to detect the data exfiltration phase of the double extortion ransomware attacks. We employ \emph{Kitsune} network feature \cite{mirsky2018kitsune}; \emph{Kitsune} is a network feature used in the \emph{Kitune} lightweight Network Intrusion Detection System (NIDS). Although the original \emph{Kitsune} NIDS consists of an ensemble of small auto-encoders (i.e., neural networks) for anomaly detection, we use only its feature structure to detect the characteristic traffic patterns in the data exfiltration phase. The challenge in this paper is to evaluate the usefulness of lightweight \emph{Kitsune} network features obtained from the hypervisor layer in detecting ransomware's data exfiltration phase.

\subsection{Contribution and organization of this paper}

This paper's contributions are as follows:

\begin{itemize}
\item The detailed attack stages, tactics, procedures, and tools of the double extortion ransomware attacks are presented.
\item A hypervisor-based data exfiltration detection method using lightweight \emph{Kitsune} network features \cite{mirsky2018kitsune} is presented.
\item The detection method is evaluated in the lab environment that simulates the scenario of the double extortion attacks using the leaked Conti ransomware playbook \cite{conti-playbook}.
\end{itemize}

The rest of the paper is organized as follows.
Section \ref{sec:attack-phase} presents the detailed attack stages of the double extortion ransomware attacks using MITRE ATT\&CK; the tools and configurations used in the data exfiltration phase are presented based on the leaked Conti playbook.
Section \ref{sec:feature-description} presents our previous work on a hypervisor-based ransomware detection method using low-level behavioral features of memory and storage access patterns. Then, we describe the novel method to detect the data exfiltration phase using \emph{Kitsune} network features.
Section \ref{sec:design-implementation} presents the design and implementation.
Section \ref{sec:evaluation} shows how much the \emph{Kitsune} network features improve the detection rate in the data exfiltration phase of double extortion attacks.
Section \ref{sec:discussion} presents the limitations and future work. We conclude the summary in Section \ref{sec:conclusion}.

\section{Attack stages of double extortion ransomware attacks}\label{sec:attack-phase}

Table \ref{tbl:att-ck} shows the mapping between attack stages of double extortion ransomware attacks and tactics of MITRE Adversarial Tactics, Techniques, and Common Knowledge (ATT\&CK). The four attack stages in the left-most column and the 10 tactics of MITRE ATT\&CK in the second left-most column were presented by Singh and Tripathy\cite{singh2024s}; we added procedures and tools in each tactic in the right-most column based on the leaked Conti playbook \cite{conti-playbook, allanliska-ransomware2023}. The leaked Conti playbook written in Cyrillic for Ransomware-as-a-Service affiliates is a manual to conduct large-scale, damaging ransomware campaigns; Ransomware-as-a-Service operators do not conduct attacks but develop and sell infrastructure and playbooks to affiliates. The affiliates do not need detailed knowledge and infrastructure for attacks. For example, the Conti playbook contains the concrete procedures, commands, and tools to obtain admin access in enterprise networks operated using Microsoft Active Directory (i.e., Kerberos-based authentication and authorization system for Windows-based enterprise networks). 

\begin{table*}[t]
\caption{Stages, tactics, procedures, and tools used in double extortion ransomware attacks; the procedures and tools are based on the leaked Conti playbook \cite{conti-playbook, allanliska-ransomware2023}.}
\label{tbl:att-ck}
\begin{center}
\begin{tabular}{llp{120mm}}
\hline
\noalign{\vskip 2.5pt}
Stage     & Tactics & Procedures and tools\\
\noalign{\vskip 2.5pt}
\hline
\noalign{\vskip 2.5pt}
Initial stage   & Reconnaissance & checks the company revenue \\
     & Initial Access & performs phishing attacks, obtains credentials from Initial Access Brokers (IABs), accesses using the internet-facing Virtual Private Network (VPN) appliances and Remote Desktop Protocol (RDP) machines\\
\noalign{\vskip 3.5pt}
Pre-operational stage & Command and Control & once the attacker intrudes into the internal network, they establish communication channels to Command and Control (C2) servers using C2 framework such as \texttt{Cobalt Strike} and \texttt{Sliver}\\
     & Privilege Escalation & obtain administrator's privilege using tools such as \texttt{Rebeus} (Kerberoasting attacks) and \texttt{Mimikatz}\\
     & Persistence & as soon as administrator rights are granted, download \texttt{AnyDesk} or \texttt{Atera} remote access application and set up it\\
     & Discovery & scans the internal local area network (LAN) using tools such as \texttt{NetScan} and examine domain information\\
\noalign{\vskip 3.5pt}
Operational stage  & Credential Access & logs on to the Domain Controller server using the administrator accounts obtained in the previous stage\\
     & Lateral Movement & logs on to other computers using obtained credentials in the previous stage\\
\noalign{\vskip 3.5pt}
Final stage   & Exfiltration & uploads confidential data using tools such as \texttt{rclone} and \texttt{MegaSync} \\
     & Impact & executes ransomware programs on the organization's entire computers to encrypt the organization's files using the compromised Domain Controller and shows ransom notes for extortion\\
\noalign{\vskip 2.5pt}
\hline 
\end{tabular}
\end{center}
\end{table*}
 
Since this paper focuses on the data exfiltration phase of the double extortion ransomware, this section presents concrete procedures and tools used in the data exfiltration phase. In the later section, we simulate the data exfiltration phase in our lab environment, where our detection method, described in section \ref{sec:feature-description} and \ref{sec:design-implementation}, is tested. The procedures and tools of the data exfiltration phase shown in the playbook \cite{conti-playbook} are as follows:

\begin{enumerate}
\item registers a new account at the Mega cloud storage (https://mega.io/).
\item downloads \texttt{rclone} program \cite{rclone} from the official site and creates a configuration file of \texttt{rclone} to sync files to the Mega cloud storage.
\item executes \texttt{rclone} program with parameters, including target directory, destination directory in the Mega cloud storage, network bandwidth limit, number of threads used to transfer, and number of file transfers in parallel. An attacker specifies a target directory in the victim organization's computers and network shares (i.e., a shared directory in an enterprise network). The exfiltrated data are sent to the Mega cloud storage.
\item copies the stolen sensitive data in the Mega cloud storage to the dedicated server using tools such as \texttt{MegaSync}.
\item accesses the Mega cloud storage using anonymizing tools such as \texttt{Tor}; then search keywords such as ``cyber insurance'' and ``corporate security policy'' in the stolen documents for later extortion purposes.
\end{enumerate}

The playbook uses \texttt{rclone} program, a legitimate cloud storage management program, for data exfiltration purposes\cite{rclone}. The \texttt{rclone} program supports many major cloud storage providers, including Google Drive and iCloud Drive, in addition to the Mega cloud storage. Fig.~\ref{fig:rclone-commands} shows an example command of \texttt{rclone} program in the leaked Conti playbook. In the playbook, the option ``\verb|--multi-thread-streams|'' (i.e., the number of threads used to transfer) is set between 1 and 12, and the option ``\verb|--transfers|'' (i.e., the number of file transfer to run in parallel) is set between 3 and 12; the authors of the playbook did not recommend the use of the maximum value (12) for the option ``\verb|--transfers|'' since the exfiltration process is more likely to be detected. The option ``\verb|--bwlimit|'' means bandwidth limits in bytes/s to upload and download files. For example, when the option ``\verb|--bwlimit 5M|'' is specified, the bandwidth is limited to 5MB/s. 

\begin{figure*}[t]
\begin{center}
\texttt{shell rclone.exe copy local-path Mega:remote-path -q --ignore-existing}
\texttt{   --auto-confirm --multi-thread-streams 1 --transfers 3 --bwlimit 5M}
\end{center}
\caption{An example of the \texttt{rclone} command for sensitive data exfiltration \cite{conti-playbook}.}
\label{fig:rclone-commands}
\end{figure*}
 
In our preliminary experiment, we counted the number of packets sent to and received from the Mega cloud storage by changing the following three parameters: ``\verb|--multi-thread-streams|'', ``\verb|--transfers|'', and ``\verb|--bwlimit|''. We specified the same values (i.e., 3, 6, 9, and 12) both for ``\verb|--multi-thread-streams|'' and ``\verb|--transfers|'' and specified four values (i.e., 0, 5M, 50M, and 100M) for ``\verb|--bwlimit|''; the bandwidth is unlimited when we specify 0 for  ``\verb|--bwlimit|''. In the experiment, we connected the test computer to the commercial optical fiber network with a bandwidth of 1 Gbps on a best-effort basis. Table \ref{tbl:spec} shows the computer's specifications used in the experiment. 

\begin{figure}[t]
\centering
\includegraphics[scale=0.38]{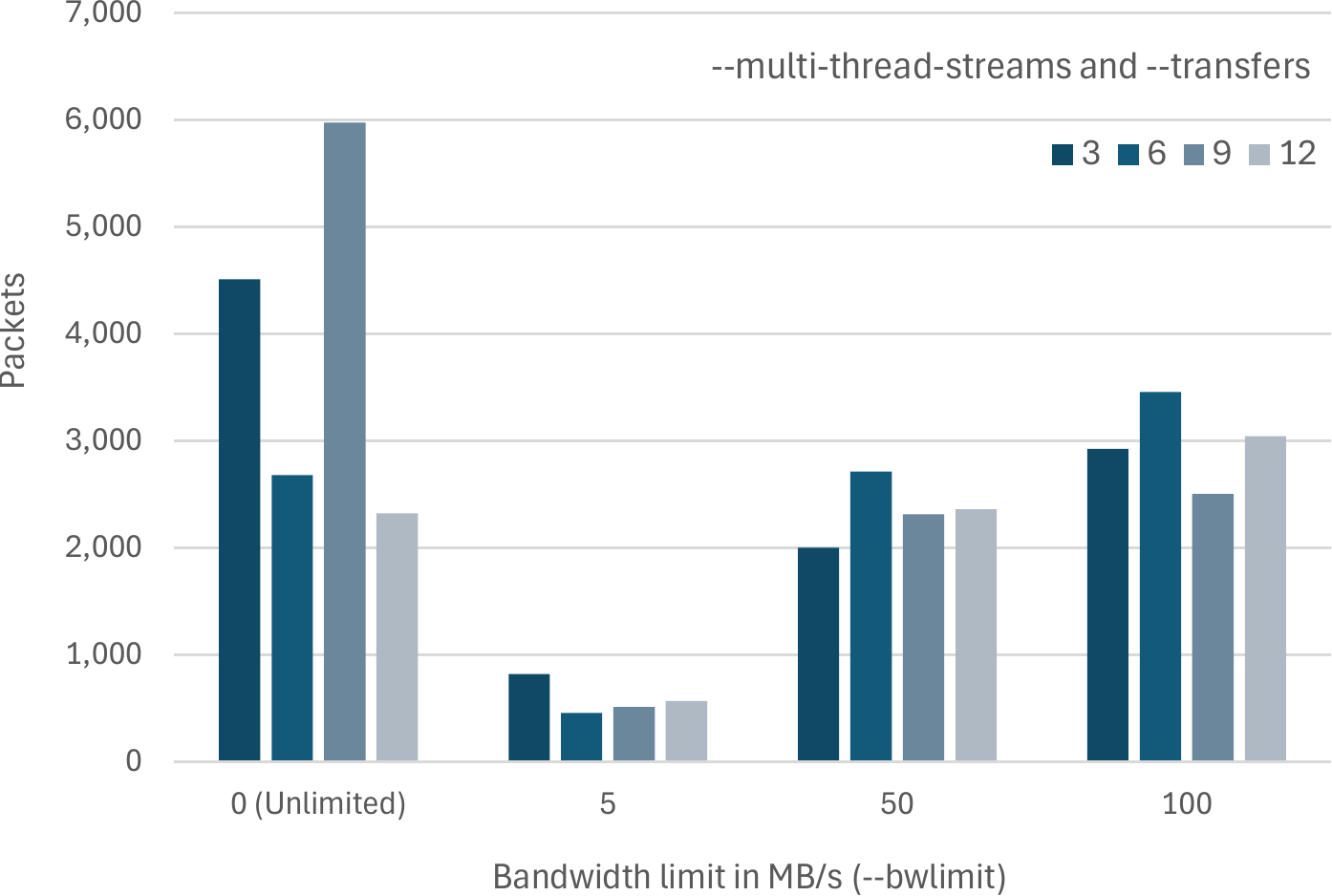}
\caption{The number of packets processed by the \texttt{rclone} program in the preliminary experiment.}
\label{fig:rclone-experiment}
\end{figure}

Fig.~\ref{fig:rclone-experiment} shows the number of packets sent and received in 30 s in the experiment. We confirmed the impact of the options of ``\verb|--multi-thread-streams|'' and ``\verb|--transfers|'' is limited compared to the option ``\verb|--bwlimit|'' from the perspective of the total number of packets (i.e., throughput of data exfiltration); therefore, we decided to employ the option ``\verb|--bwlimit|'' to evaluate our data extortion ransomware detection method in the later section. In addition, since we used a commercial Internet service provider and cloud storage provider, the throughput fluctuations were difficult to avoid, especially when we set the option ``\verb|--bwlimit 0|'' that means no bandwidth limit; we thus confirmed that the detection method using network features have to handle some throughput fluctuations to adapt various network environments. 
In the next section, we present a detection method for the data exfiltration phase and encryption phase of ransomware attacks. 

\begin{table}[t]
\caption{Specification of the test machine used in developing hypervisor-based double extortion detection method.}
\label{tbl:spec}
\begin{center}
\begin{tabular}{p{12mm}l}
\hline
\noalign{\vskip 1.0pt}
CPU & Intel Core i3 12100 (4 P-cores, 8 threads) \\
RAM & DDR4 2133 8GiB x 2 \\
Motherboard & ASRock B660M-HDV \\
SSD & Crucial CT240BX, CT250MX, Samsung 840 \\
Network & Intel Pro1000 NIC (1 Gbps), Intel X550-T1 NIC (10 Gbps) \\
Hypervisor & BitVisor downloaded on 25th Jan. 2022 (8c129a1) \\
Guest OS & Windows 10 LTSC \\
\noalign{\vskip 0.5pt}
\hline 
\end{tabular}
\end{center}
\end{table}

\section{Low-level memory and storage behavioral features and Kitsue network features for double extortion ransomware detection}
\label{sec:feature-description}

We presented a hypervisor-based monitoring system to collect low-level behavioral features for detecting ransomware's destruction phase (i.e., \emph{impact} in MITRE ATT\&CK)\cite{HIRANO2022301314, HIRANO2025104202}; the thin hypervisor works as an additional protection layer to the conventional OS-level ransomware protection layer. We briefly describe the previous work.

\subsection{Thin-hypervisor-based monitoring system}

\begin{figure}[tb]
\centering
\includegraphics[scale=0.65]{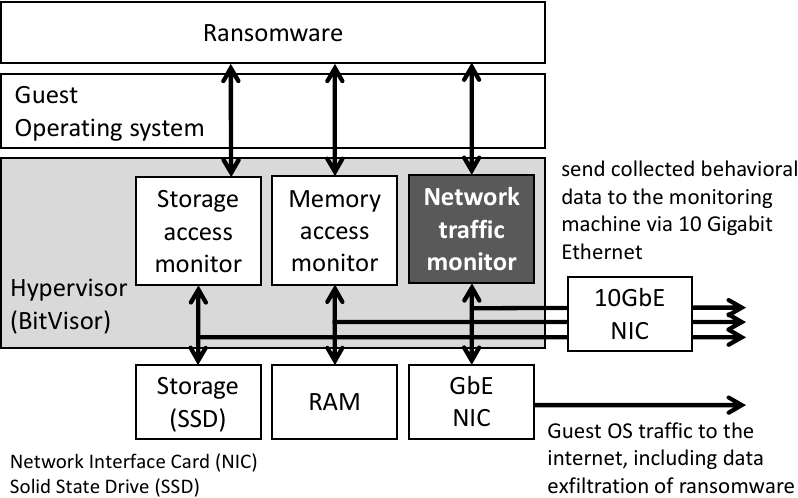}
\caption{Hypervisor-based monitoring system to collect low-level behavioral data \cite{HIRANO2022301314,HIRANO2025104202}; the network traffic monitor is a novel function to detect data exfiltration phase of double extortion ransomware attacks.}
\label{fig:monitoring-system}
\end{figure}

Fig.~\ref{fig:monitoring-system} shows the hypervisor-based monitoring system developed using a thin hypervisor named BitVisor \cite{bitvisor}; storage access monitor collects access patterns on storage devices (e.g., Solid State Drive) using an extended Advanced Host Controller Interface (AHCI) para-pass-through driver of the hypervisor \cite{HIRANO2022301314}. The memory access patterns on RAM are collected using a hardware-assisted memory virtualization technology named Intel's Extended Page Table (EPT) \cite{HIRANO2025104202, intel-sdm}. Our previous paper \cite{HIRANO2025104202} showed that a deep-learning-based ransomware detector trained using the low-level storage and memory behavioral features collected using the thin-hypervisor-based monitoring system can detect crypto-ransomware in 0.964 of F score.

In this paper, we developed a novel network traffic monitor in the thin hypervisor software (Fig.~\ref{fig:monitoring-system}). We added the network monitoring function to BitVisor's para-pass-through Network Interface Card (NIC) driver of Intel Pro 1000 (pro1000.c). The network traffic monitor collects all Ethernet frames sent and received through the Gigabit Ethernet Network Interface Card (GbE NIC) the guest OS uses. The obtained data consists of 1,518-byte Ethernet frames. The Ethernet frames, in addition to memory and storage access pattern data, collected in the thin hypervisor are sent to the monitoring machine via a dedicated 10 Gigabit Ethernet Network Interface Card (10GbE NIC); the 10GbE NIC is concealed from the guest OS by the thin hypervisor to prevent attacks against the monitoring function. We use the received data on the monitoring machine to train deep learning models.

\subsection{Behavioral features to train deep learning models for ransomware detection}

\begin{figure}[tb]
\centering
\includegraphics[scale=0.42]{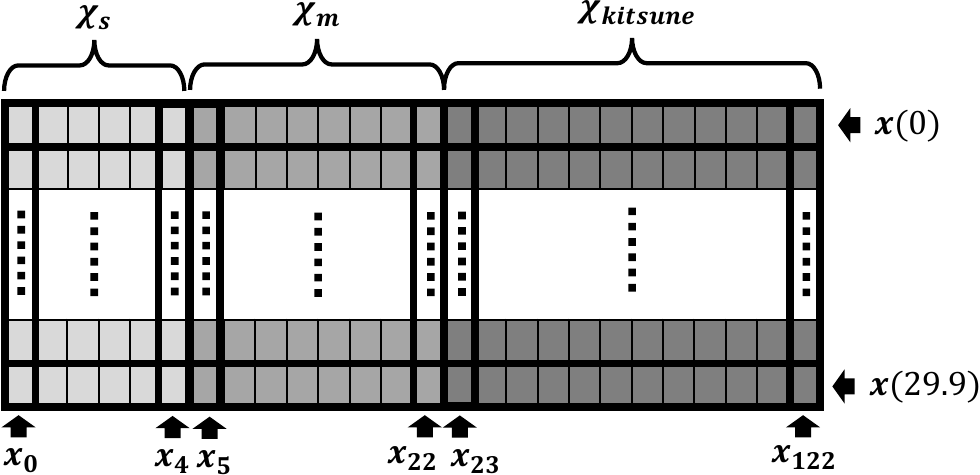}
\caption{Structure of a behavioral feature $\bm{\chi}$ consisting of $\bm{\chi_s}$, $\bm{\chi_m}$, $\bm{\chi_{kitsune}}$ in 30 s after executing a malicious program or benign program ($T_{window}$\,=\,0.1 s and $T_d$\,=\,30 s).}
\label{fig:feature-description}
\end{figure}

Fig.~\ref{fig:feature-description} shows a structure of a behavioral feature $\bm{\chi}$ that consists of a five-dimensional storage feature $\bm{\chi_s}=\{x_1,\dots,x_4\}$, a 18-dimensional memory feature $\bm{\chi_m}=\{x_5,\dots,x_{22}\}$, and a 100-dimensional \emph{Kitsune} network feature $\bm{\chi_{kitsune}}=\{x_{23},\dots, x_{122}\}$; a \emph{Kitsune} network feature $\bm{\chi_{kitsune}}$ is a new feature introduced in this paper. Thus, the 123-dimensional feature $\bm{\chi}$ is created to train deep-learning models for ransomware detection. 
Fig.~\ref{fig:feature-description} presents the 123-dimensional feature $\bm{\chi}$ in 30 s after executing a malicious program (i.e., a ransomware executable and a data exfiltration program) or a benign program. The 123-dimensional feature $\bm{\chi}$ is created using $T_{window}$\,=\,0.1 s and $T_{d}$\,=\,30; $T_{window}$ is a period to calculate a feature vector $\bm{x}$. $T_d$ is the duration of access patterns used in detecting ransomware; therefore, $T_{d}$ is also referred to as the detection time of ransomware. 
Fig.~\ref{fig:feature-description} shows the feature structure of $\bm{\chi}=\{\bm{\chi_s}, \bm{\chi_m}, \bm{\chi_{kitsune}}\}=\{\bm{x}(0), \bm{x}(0.1), \bm{x}(0.2),..., \bm{x}(29.9)\}$. Each row represents a feature vector at a specific time window; for example, $\bm{x}(0)$ is a 123-dimensional feature vector between 0 s and 0.1 s. On the other hand, a behavioral feature $\bm{\chi}$ can also be expressed in $\bm{\chi}=\{\bm{\chi_s}, \bm{\chi_m}, \bm{\chi_{kitsune}} \}=\{ \bm{x_0}, \bm{x_1}, \bm{x_2},..., \bm{x_{122}}\}$. Each column represents a specific feature vector between 0 s and 30 s. For example, $\bm{x_{23}}=\{ \bm{x_{23}}(0), \bm{x_{23}}(0.1), \bm{x_{23}}(0.2),..., \bm{x_{23}}(29.9) \}$ is a feature vector of the first element of \emph{Kitsune} network feature $\bm{\chi_{kitsune}}$ between 0 s and 30 s. $\bm{x_i}(j)$ is a $i$th feature vector calculated using storage and memory access patterns and network traffic between $j$ s and $(j + T_{window})$ s.

\subsection{\emph{Kitsune} network features to detect data exfiltration}

Mirsky et al. presented a lightweight \emph{Kitsune} Network Intrusion Detection System (NIDS) designed to deploy on inexpensive distributed routers \cite{mirsky2018kitsune}; \emph{Kitsune} means fox in Japanese and foxes are famous for tricking humans by changing their appearance except for their tails in Japanese folklore. The original \emph{Kitsune} NIDS detects anomaly using ensembles of small auto-encoders (i.e., neural networks) trained using \emph{Kitsune} network feature $\bm{\chi_{kitsune}}$. In this paper, we employ only the feature extractor of \emph{Kitsune} NIDS and do not use ensembles of auto-encoders. 

\emph{Kitsune} NIDS support high-speed feature extraction of temporal statistics over a dynamic number of network channels with a small memory footprint; the features are extracted using \emph{incremental statistics}, which means feature vectors that are updated using transmitted or received packet size with elapsed-time-based decay factors. 
Let $S=\{x_1, x_2, \dots\}$ be a sequence of observed packet sizes where $x_i\in\mathbb{R}$. A tuple $IS:=(N, LS, SS)$ consists of the number of received packets, a linear sum, and a squared sum of packet sizes transmitted or received so far. The update procedure of $IS$ for the packet size $x_i$ is $IS\leftarrow(N+1, LS+x_i, SS+x_i^2)$. The \emph{Kitsune} NIDS uses five time-window of 100 ms, 500 ms, 1.5 s, 10 s, and 1 min to attenuate \emph{incremental statistics} $IS$; values in $IS$ of each time window are attenuated by applying each of decay factors $\lambda = \{$5, 3, 1, 0.1, 0.01\}; the decay factors were chosen in the original paper \cite{mirsky2018kitsune}.

Table \ref{tbl:kitsune-1d-2d} shows two types of \emph{incremental statistics} used in \emph{Kitsune} NIDS. A one-dimensional (1D) \emph{incremental statistics} consists of three real numbers of statistics (i.e., $\mathbb{R}^3$). On the other hand, a two-dimensional (2D) \emph{incremental statistics} consists of four real numbers of statistics (i.e., $\mathbb{R}^4$). The values of ID and 2D \emph{incremental statistics} are maintained in a hash table; a key-value pair in the hash table is maintained using the following keys:

\begin{itemize}
\item \textbf{srcMAC+srcIP}: A string that concatenates a source Media Access Control (MAC) address and source Internet Protocol (IP) address
\item \textbf{srcIP+dstIP}: A string that concatenates a source IP address and destination IP address
\item \textbf{srcIP, dstIP}: A source IP address and destination IP address
\item \textbf{srcIP+srcPort, dstIP+dstPort}: A string that concatenates a source IP address and source port number and a string that concatenates a destination IP address and destination port number
\end{itemize}

An 1D \emph{incremental statistics} is calcurated for each of the keys in \textbf{srcMAC+srcIP}, \textbf{srcIP+dstIP}, \textbf{srcIP}, and \textbf{srcIP+srcPort}. A 2D \emph{incremental statistics} is calculated for each pair of the keys \textbf{(srcIP, dstIP)} and \textbf{(srcIP+srcPort, dstIP+dstPort)}; the 2D statistics represent dual traffic behavior between source and destination. Therefore, a \emph{Kitsune} network feature in a single time window contains four 1D incremental statistics consisting of 12 real numbers (i.e., $\mathbb{R}^{12}$) and two 2D incremental statistics consisting of eight real numbers (i.e., $\mathbb{R}^8$). We concatenate the \emph{Kitsune} network feature of five time windows; thus, we obtains a 100-dimensional \emph{Kitsune} network feature $\bm{\chi_{kitsune}}$ (i.e., $\mathbb{R}^{100}$) in total. We used the 100-dimensional \emph{Kitsune} network features $\bm{\chi_{kitsune}}$ of the last packet processed in each time window $T_{window}$.

\begin{table}
\caption{1D and 2D incremental statistics used in \emph{Kitsune} NIDS.}
\label{tbl:kitsune-1d-2d}
\begin{center}
\begin{tabular}{cccp{35mm}}
\hline
\noalign{\vskip 2.5pt}
Type & Statistic & Notation & Calculation \\
\noalign{\vskip 2.5pt}
\hline
\noalign{\vskip 2.5pt}
\multirow{4}{*}{1D} & Weight & $w$ & $w$ \\
\noalign{\vskip 2.5pt}
     & Mean & $\mu_{S_i}$ & $LS/w$ \\
\noalign{\vskip 2.5pt}
     & Stdard & \multirow{2}{*}{$\sigma_{S_i}$} & \multirow{2}{*}{$\sqrt{|SS/w - (LS/w)^2|}$} \\
     & Deviation & & \\
\noalign{\vskip 2.5pt}
\hline
\noalign{\vskip 2.5pt}
\multirow{6}{*}{2D} & Magnitude & $\|S_i,S_j\|$ & $\sqrt{{\mu_{S_i}}^2+{\mu_{S_j}}^2}$ \\
\noalign{\vskip 2.5pt}
     & Radius  & $R_{S_i,S_j}$ & $\sqrt{({\sigma_{S_i}}^2)^2+({\sigma_{S_j}}^2)^2}$ \\
\noalign{\vskip 4.0pt}
     & Approx.  & \multirow{2}{*}{ $Cov_{S_i,S_j}$ } & \multirow{2}{*}{ $\dfrac{SR_{ij}}{w_i+w_j}$ } \\
     & Covariance & & \\
\noalign{\vskip 5.0pt}
     & Correlation & \multirow{2}{*}{ $P_{S_i,S_j}$ } & \multirow{2}{*}{ $\dfrac{Cov_{S_i,S_j}}{\sigma_{S_i}\sigma_{S_j}}$ } \\
     & Coefficient & & \\
\noalign{\vskip 5.0pt}
\hline 
\end{tabular}
\end{center}
\end{table}

\section{Design and implementation}
\label{sec:design-implementation}

\begin{figure}[tb]
\centering
\includegraphics[scale=0.7]{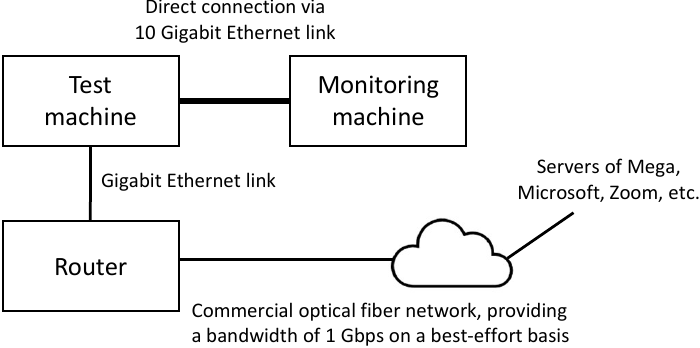}
\caption{A lab environment to test the detection system.}
\label{fig:test-setup}
\end{figure}

We developed the network traffic monitor function in Gigabit Ethernet Network Interface Card (NIC) para-pass-through driver of the thin hypervisor (BitVisor) described in Section \ref{sec:feature-description}. Table~\ref{tbl:spec} shows the test machine's specification in developing the hypervisor-based double extortion ransomware detection method.
Fig.~\ref{fig:test-setup} shows a lab environment to test the detection system. We connected the test machine, where the hypervisor-based monitoring system runs, to the Internet using a commercial optical fiber network with a bandwidth of 1 Gbps on a best-effort basis. The test machine that executes the thin-hypervisor-based monitoring system is connected to the monitoring machine via a direct 10 Gigabit Ethernet link; the monitoring machine receives the collected storage and memory access patterns and Ethernet frames sent to and received from the Internet on the test machine.

In the evaluation, we collected behavioral feature $\bm{\chi}$ of Firefox with autopilot plugin, Microsoft 365 Office applications (i.e., Excel, PowerPoint, Word) that open a remote file from OneDrive and process them using a macro program or play slide show, teleconference applications (i.e., Microsoft Teams and Zoom) that play YouTube video of Mirsky's \emph{Kitsune} NIDS presentation recorded at NDSS2018 conference in a meeting with two participants, and \texttt{MegaSync} and \texttt{rclone} programs that sent dummy files to cloud storage. We copied the first 50 directories of the \emph{GovDocs1} dataset \cite{govdocs} to the Desktop on the test machine. The \texttt{rclone} program is configured to send dummy files to Google Drive or Mega; we configured the parameter of ``\verb|--bwlimit|' to 5 MB/s, 50 MB/s, and 100 MB/s. We executed each program 10 times to create the dataset.
We used the same deep neural network model consisting of a one-dimensional Convolutional Neural Network (1D-CNN) layer and two Long Short-Term Memory (LSTM) layers used in the previous paper \cite{HIRANO2025104202}, trained and evaluated the model using the obtained feature $\bm{\chi}$. We used 70\% of the dataset for training and 30\% of the dataset for prediction.

\section{Evaluation}
\label{sec:evaluation}

\begin{figure*}[tb]
\centering
\includegraphics[scale=0.7]{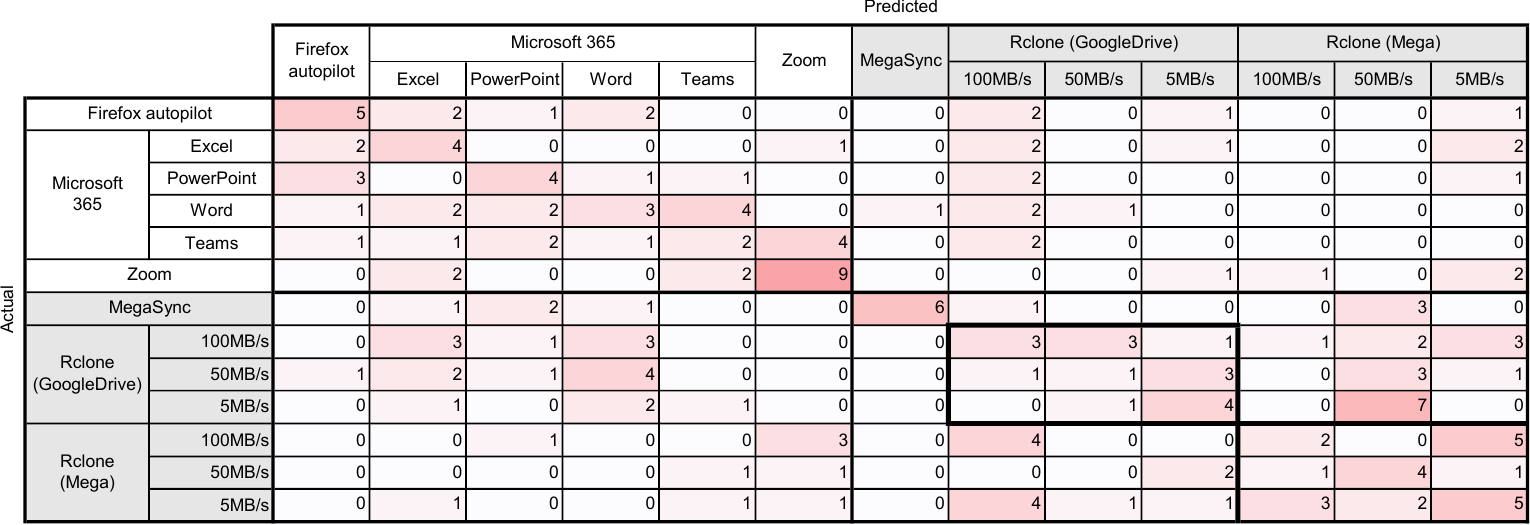}
\caption{Confusion matrix of deep learning model trained using only storage and memory feature $\bm{\chi}=\{\bm{\chi_{s}},\bm{\chi_{m}}\}$.}
\label{fig:13classes-atamem}
\end{figure*}

We evaluated the impact of introducing \emph{Kitsune} network feature $\bm{\chi_{kitune}}$ in addition to storage behavioral feature $\bm{\chi_{s}}$ and memory behavioral feature $\bm{\chi_{m}}$.
Fig.~\ref{fig:13classes-atamem} shows the confusion matrix of the deep learning model trained using only storage and memory access patterns $\bm{\chi}=\{\bm{\chi_{s}},\bm{\chi_{m}}\}$. The confusion matrix was created using the sum of the five model predictions. The average micro F-score of 13 classifications in five trials was 0.267 when we used only storage and memory features.

Fig.~\ref{fig:13classes-atamem} shows the confusion matrix of the deep learning model trained using Kitsune network features in addition to storage and memory access patterns $\bm{\chi}=\{\bm{\chi_{s}},\bm{\chi_{m}},\bm{\chi_{kitsune}}\}$. The confusion matrix was created using the sum of the five model predictions. The average micro F-score of 13 classifications was 0.513 in five trials when we used \emph{Kitsune} network feature in addition to storage and memory features. We improved 0.246 in the micro F score of 13 classifications using \emph{Kitsune} network feature.

Next, we assume that \texttt{MegaSync} and \texttt{rclone} programs belong to a malicious class that exfiltrates sensitive data; other programs (i.e., Firefox, three Microsoft Office applications, Teams, and Zoom) belong to a benign class. The macro F score of the binary classification was 0.721 when we used only storage and memory features. On the other hand, the macro F score of the binary classification was 0.887 when we used \emph{Kitsune} network feature in addition to storage and memory features. We thus confirmed that the presented method improved by 0.166 in the macro F score of binary classification using \emph{Kitsune} network feature.

\begin{figure*}[tb]
\centering
\includegraphics[scale=0.7]{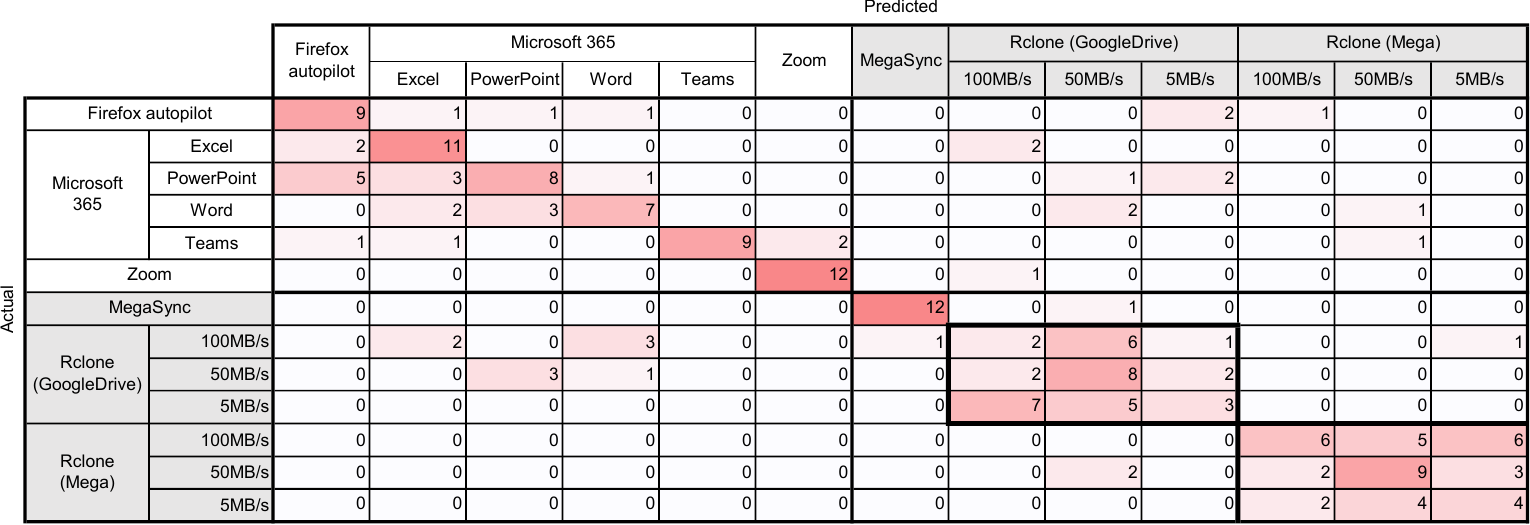}
\caption{Confusion matrix of deep learning model trained using \emph{Kitsune} network features in addition to storage and memory features $\bm{\chi}=\{\bm{\chi_{s}},\bm{\chi_{m}},\bm{\chi_{kitsune}}\}$.}
\label{fig:13classes-all}
\end{figure*}

\section{Discussion}\label{sec:discussion}

We first discuss the experimental results in detail. In the experiment, we used three bandwidth limits (i.e., 5 MB/s, 50 MB/s, and 100 MB/s) and two cloud storage providers (i.e., Google Drive and Mega) for the \texttt{rclone} program. Fig.~\ref{fig:13classes-all} shows that the developed detection system could distinguish the two cloud storage providers; however, it was difficult to determine the differences in bandwidth limits. In addition, the developed detection system could distinguish the behaviors between \texttt{MegaSync} and \text{rclone} programs that use the Mega cloud storage. We also confirmed that the developed system could distinguish the behaviors between Office applications with low-traffic patterns (i.e., Excel, PowerPoint, and Word). The detection rates of two teleconference applications (i.e., Teams and Zoom) also improved.

The limitations and future work of the presented detection method are as follows: (1) although the current system shown in Fig.~\ref{fig:monitoring-system} has the monitoring function of storage and memory access patterns and network traffic, it does not have the deep-learning-based ransomware detection function in the thin hypervisor; we created the model and predicted results in another machine for now. Therefore, our future work includes implementing machine-learning or deep-learning model update and prediction functions in the hypervisor software. (2) this paper examined only the data exfiltration phase using the limited tools shown in the leaked Conti playbook\cite{conti-playbook}; we need to examine earlier stages of the double extortion ransomware attacks, including lateral movement, credential access, discovery, persistence, privilege escalation, command and control, and initial access of MITRE ATT\&CK tactics described in Section \ref{sec:attack-phase}. The early detection will reduce the security risk of the victim organizations. (3) since the presented method uses a hypervisor to collect storage and memory access patterns and network traffic, it cannot distinguish the behaviors of multiple applications executed in parallel, for example, including the operating system's processes and benign high-bandwidth transfers (e.g., large backups). Although we conducted the experiment executing a single application at a time, we need further realistic conditions that run multiple applications simultaneously. Therefore, the challenges in our future work include mitigating some degree of semantic gap problem between the operating system and hypervisor (e.g., distinguishing each process in hypervisor software). Tapaswi defined the semantic gap problems in Virtual Machine Introspection (VMI) techniques as difficulties in deriving a complete view of the guest operating system from the hypervisor due to the highly dynamic nature of modern operating systems and virtualization software \cite{more2014virtual}.
Many ransomware detection methods using network activity have been implemented on Windows but not on hypervisors, unlike our method \cite{cen2024ransomware}.


\section{Conclusion}\label{sec:conclusion}

As organizations prepare for crypto-ransomware attacks using more robust backup strategies, more ransomware attackers have focused on sensitive data exfiltration. The early detection of such double extortion ransomware attacks is crucial to keep the organization's security risk low. This paper first presented the detailed attack stages, tactics of MITRE ATT\&CK, procedures, and tools used in the real double-extortion ransomware attacks based on the leaked Conti playbook\cite{conti-playbook}. We developed the novel network traffic monitor function in the thin hypervisor of our previous work \cite{HIRANO2025104202}; we employed lightweight network feature used in \emph{Kitsune} Network Intrusion Detection System (NIDS) in training our deep-learning-based ransomware detector. 

We examined the two data exfiltration tools (i.e., \texttt{rclone} and \texttt{MegaSync} programs) and six benign applications, including a web browser with autopilot plugin, three Office applications that open and manipulate files on OneDrive, and two teleconference applications. The presented method improved 0.166 in the macro F score of binary classification (i.e., data exfiltration programs and benign programs) compared to the method that used only storage and memory features. We finally discussed the limitations of the current work, including the lack of detection function in the hypervisor software, the need to detect earlier attack stages, and examining detection performance in more realistic conditions of simultaneous execution of applications that need to solve some semantic gap problems.

\section*{Acknowledgment}

This work was supported by JSPS KAKENHI Grant Number JP23K11114. The authors gratefully acknowledge constructive comments by the anonymous reviewers. The authors thank Yuta Fukino, Tomohiro Yamaguchi, Eitaro Osugi for their support in constructing and analyzing the dataset. The authors gratefully thank the developers of BitVisor\cite{bitvisor}.

\bibliographystyle{IEEEtran}
\bibliography{hirano}

\begin{thebibliography}{10}
\providecommand{\url}[1]{#1}
\csname url@samestyle\endcsname
\providecommand{\newblock}{\relax}
\providecommand{\bibinfo}[2]{#2}
\providecommand{\BIBentrySTDinterwordspacing}{\spaceskip=0pt\relax}
\providecommand{\BIBentryALTinterwordstretchfactor}{4}
\providecommand{\BIBentryALTinterwordspacing}{\spaceskip=\fontdimen2\font plus
\BIBentryALTinterwordstretchfactor\fontdimen3\font minus
  \fontdimen4\font\relax}
\providecommand{\BIBforeignlanguage}[2]{{%
\expandafter\ifx\csname l@#1\endcsname\relax
\typeout{** WARNING: IEEEtran.bst: No hyphenation pattern has been}%
\typeout{** loaded for the language `#1'. Using the pattern for}%
\typeout{** the default language instead.}%
\else
\language=\csname l@#1\endcsname
\fi
#2}}
\providecommand{\BIBdecl}{\relax}
\BIBdecl

\bibitem{mcintosh2021ransomware}
T.~McIntosh, A.~Kayes, Y.-P.~P. Chen, A.~Ng, and P.~Watters, ``Ransomware
  mitigation in the modern era: A comprehensive review, research challenges,
  and future directions,'' \emph{ACM Computing Surveys (CSUR)}, vol.~54, no.~9,
  pp. 1--36, 2021.

\bibitem{conti-playbook}
W.~Largent, ``{Translated: Talos' insights from the recently leaked Conti
  ransomware playbook},''
  \url{https://blog.talosintelligence.com/conti-leak-translation/}, 2021,
  accessed 4 April 2025.

\bibitem{begovic2023cryptographic}
K.~Begovic, A.~Al-Ali, and Q.~Malluhi, ``Cryptographic ransomware encryption
  detection: Survey,'' \emph{Computers \& Security}, vol. 132, p. 103349, 2023.

\bibitem{cen2024ransomware}
M.~Cen, F.~Jiang, X.~Qin, Q.~Jiang, and R.~Doss, ``Ransomware early detection:
  A survey,'' \emph{Computer Networks}, vol. 239, p. 110138, 2024.

\bibitem{singh2024s}
N.~Singh and S.~Tripathy, ``It's too late if exfiltrate: Early stage android
  ransomware detection,'' \emph{Computers \& Security}, vol. 141, p. 103819,
  2024.

\bibitem{oz2022survey}
H.~Oz, A.~Aris, A.~Levi, and A.~S. Uluagac, ``A survey on ransomware:
  Evolution, taxonomy, and defense solutions,'' \emph{ACM Computing Surveys
  (CSUR)}, vol.~54, no. 11s, pp. 1--37, 2022.

\bibitem{HIRANO-CSR2022}
M.~Hirano and R.~Kobayashi, ``Machine learning-based ransomware detection using
  low-level memory access patterns obtained from live-forensic hypervisor,'' in
  \emph{2022 IEEE International Conference on Cyber Security and Resilience
  (CSR)}, 2022, pp. 323--330.

\bibitem{HIRANO2022301314}
M.~Hirano, R.~Hodota, and R.~Kobayashi, ``{RanSAP: an open dataset of
  ransomware storage access patterns for training machine learning models},''
  \emph{Forensic Science International: Digital Investigation}, vol.~40, p.
  301314, 2022.

\bibitem{HIRANO2025104202}
M.~Hirano and R.~Kobayashi, ``{RanSMAP: Open dataset of Ransomware Storage and
  Memory Access Patterns for creating deep learning based ransomware
  detectors},'' \emph{{Computers \& Security}}, vol. 150, 2025, 104202.

\bibitem{mirsky2018kitsune}
Y.~Mirsky, T.~Doitshman, Y.~Elovici, and A.~Shabtai, ``Kitsune: an ensemble of
  autoencoders for online network intrusion detection,'' \emph{arXiv preprint
  arXiv:1802.09089}, 2018.

\bibitem{allanliska-ransomware2023}
A.~Liska, \emph{{Ransomware: Understand. Prevent. Recover. 2nd Edition}}.\hskip
  1em plus 0.5em minus 0.4em\relax {ActualTech Media}, 2023.

\bibitem{rclone}
N.~Craig-Wood, ``{Rclone},'' \url{https://rclone.org/}, 2014, accessed 12 April
  2025.

\bibitem{bitvisor}
T.~Shinagawa, H.~Eiraku, K.~Tanimoto, K.~Omote, S.~Hasegawa, T.~Horie,
  M.~Hirano, K.~Kourai, Y.~Oyama, E.~Kawai \emph{et~al.}, ``{BitVisor: A Thin
  Hypervisor for Enforcing I/O Device Security},'' in \emph{{Proceedings of the
  2009 ACM SIGPLAN/SIGOPS international conference on Virtual execution
  environments (VEE 2009)}}, 2009, pp. 121--130.

\bibitem{intel-sdm}
{Intel Corporation}, ``{Intel{\textregistered} 64 and IA-32 architectures
  software developer’s manual},'' \emph{{Volume 3C: system programming guide,
  part 3}}, 2021.

\bibitem{govdocs}
S.~L. Garfinkel, ``{GovDocs1 - Digital Corpora},''
  \url{https://digitalcorpora.org/corpora/files}, accessed 16 April 2022.

\bibitem{more2014virtual}
A.~More and S.~Tapaswi, ``Virtual machine introspection: towards bridging the
  semantic gap,'' \emph{Journal of Cloud Computing}, vol.~3, no.~1, pp. 1--14,
  2014.

\end{thebibliography}

\end{document}